\documentclass[
	aps, prd, reprint,
	10pt, notitlepage, a4paper,
         floats, floatfix,
	amsmath, amssymb, amsfonts, eqsecnum,
	superscriptaddress,
	showpacs, showkeys,
	nofootinbib,
 	longbibliography,
]{revtex4-1}

\usepackage[usenames,dvipsnames]{xcolor}
\colorlet{LightRubineRed}{RubineRed!70}
\colorlet{Mycolor1}{green!10!orange}
\definecolor{Mycolor2}{HTML}{00F9DE}
\usepackage{amssymb}
\usepackage{amsmath}
\usepackage{cancel}
\usepackage{tikz}
\usepackage{caption}
\usepackage[compat=1.1.0]{tikz-feynman}
\usetikzlibrary[shapes.misc] 
\usetikzlibrary{arrows,snakes,backgrounds}
\usepackage{aas_macros}
\usepackage{todonotes}
\usepackage{xspace} 
\usepackage{bm,graphicx} 
\usepackage[utf8]{inputenc} 
\newcommand{\be}{\begin{equation}}
\newcommand{\ee}{\end{equation}}
\newcommand{\bea}{\begin{eqnarray}}
\newcommand{\eea}{\end{eqnarray}}

\newcommand{\bdm}{\begin{displaymath}}
\newcommand{\edm}{\end{displaymath}}

\xdefinecolor{mylinkcolor}{rgb}{0,0,0.5}
\usepackage[
	bookmarksnumbered, bookmarksopen, bookmarksopenlevel=2,
	breaklinks=true,
	colorlinks=true, filecolor=mylinkcolor, citecolor=mylinkcolor,
	linkcolor=mylinkcolor, urlcolor=mylinkcolor, menucolor=mylinkcolor,
]{hyperref}

\begin{document}

\title{Binary dynamics from Einstein-Maxwell theory at second post-Newtonian order using effective field theory}

\author{Pawan Kumar Gupta}\email{p.gupta@nikhef.nl}
\affiliation{Institute for Gravitational and Subatomic Physics (GRASP),\\ Department of Physics, Utrecht University, Princetonplein 1, 3584 CC Utrecht, The Netherlands, EU}
\affiliation{Nikhef, Science Park, 1098 XG Amsterdam, The Netherlands, EU}
\begin{abstract}
 The detection of gravitational waves from binary black holes sources has opened the possibility to search for electric charges and ``dark'' charges on black holes, the latter being candidates for dark matter. This requires theoretical predictions about the effects of these charges on the inspiral of binary black holes in order to place constraints on them. The effects of these charges on the inspiral can be described using Einstein-Maxwell theory. These effects have previously been derived up to the first post-Newtonian (1PN) order, and the results were recently used to place bounds on the charge-to-mass ratio of black holes. In this work, we employ the effective field theory approach, with a metric parameterization based on a temporal Kaluza-Klein decomposition and non-relativistic gravitational fields, to derive the Lagrangian for binary motion under the influence of charges up to 2PN order. Our results provide the foundation for deriving precision predictions of the effect of charges on the inspiral of binary black holes for gravitational-wave astronomy.
\end{abstract}
\maketitle


\section{Introduction}
To date, the Advanced LIGO \cite{LIGOScientific:2014pky} and Advanced Virgo \cite{VIRGO:2014yos} gravitational wave (GW) detectors have discovered nearly 90 candidate coalescing binary black hole signals \cite{LIGOScientific:2016dsl,LIGOScientific:2016aoc,LIGOScientific:2018mvr,LIGOScientific:2020ibl,LIGOScientific:2021djp}. This has opened up avenues to test various physical scenarios, particularly the possibility that binary black holes carry electric charges or dark charges associated with dark matter candidates \cite{Gupta:2021rod}. Signals from binary black holes consist of three phases: inspiral, merger, and ringdown. Charges affect the ringdown physics \cite{Pani:2013ija,Pani:2013wsa,Zilhao:2014wqa,Mark:2014aja,Dias:2015wqa,Dias:2021yju}, and in \cite{Carullo:2021oxn}, an analysis of quasi-normal modes was performed on GW150914, leading to 90\% upper bounds on the charge-to-mass ratio of the remnant black hole of $|q/m| < 0.33$. Here, we focus on the inspiral part of the signal, which can be described analytically through the post-Newtonian (PN) approximation \cite{Blanchet:2009ggi, Blanchet:2013haa}. When charges are introduced, the appropriate model is Einstein-Maxwell theory, where the Lagrangian for binary motion was previously calculated up to first PN (1PN) order in \cite{Patil:2020dme, Khalil:2018aaj} (the latter also included a dilaton). The results were subsequently used to constrain charges on coalescing binary black holes using gravitational wave signals from the second Gravitational Wave Transient Catalog (GWTC-2); in \cite{Gupta:2021rod}, 1-$\sigma$ constraints of $|q/m| \leq 0.2 - 0.3$ were placed.

Considering further improvements in detector sensitivities, particularly with the planned next-generation GW observatories Einstein Telescope \cite{Punturo:2010zza,Hild:2010id}, Cosmic Explorer \cite{LIGOScientific:2016wof}, and LISA \cite{2017arXiv170200786A,Baker:2019nia,Amaro-Seoane:2022rxf}, more accurate theoretical predictions are in order. In this work, we calculate the Lagrangian for two inspiraling bodies with charges included up to 2PN order, using the effective field theory (EFT) approach.

The EFT approach was first suggested by Goldberger and Rothstein to solve the binary inspiral problem in an alternative way \cite{Goldberger:2004jt,Goldberger:2007hy}. This framework is commonly known as NRGR (Non-Relativistic General Relativity), which we extend here to the Einstein–Maxwell system. It efficiently uses standard tools from quantum field theory, such as Feynman diagrams and dimensional regularization, but applies them to the classical two-body problem in general relativity. Currently, the framework has been used up to 5PN order \cite{Blumlein:2021txe,Kim:2021rfj,Foffa:2020nqe}, including spin-orbit and spin-spin effects up to 3PN order \cite{Porto:2005ac,Porto:2006bt,Porto:2008jj,Porto:2008tb}

A novel method within the EFT approach was suggested by Kol and Smolkin (KS), who used a temporal Kaluza-Klein parameterization of the metric to improve computational efficiency \cite{Kol:2007bc}. This non-relativistic parameterization defines a set of new non-relativistic gravitational (NRG) fields. As we shall see, when charges are included, this leads to a clear coupling hierarchy between mass and charge, making the technique particularly suited to our problem. More generally, the KS parameterization has been shown to reduce the complexity involved in calculating the Einstein-Infeld-Hoffmann (1PN) Lagrangian \cite{Kol:2007bc} and the 2PN Lagrangian \cite{Gilmore:2008gq}. Similar simplifications were observed in the calculation of various contributions to the spin-spin potential \cite{Levi:2008nh,Levi:2011eq,Levi:2014sba,Levi:2015ixa} as well as the spin-orbit potential \cite{Levi:2010zu,Levi:2015uxa,Levi:2020kvb}.

In this paper, we calculate the Lagrangian for the Einstein-Maxwell action up to 2PN order using EFT with NRG fields. We recover the Coulomb potential using a single diagram and reproduce the 1PN order Lagrangian, which requires computing only four diagrams and agrees with the known results \cite{Patil:2020dme, Khalil:2018aaj}. At 1PN order, we encounter a one-loop integral, which we solve using a master formula (Eq.~\ref{eq:1loop}). At 2PN order, we find 16 diagrams contributing at one-loop and 21 diagrams contributing at two-loop. At two-loop, we encounter integrals that diverge, which we handle using standard dimensional regularization techniques. We demonstrate how the EFT with NRG fields can be used to efficiently compute conservative binary dynamics with charge at high precision in the PN expansion.

This paper is organized as follows. In Sec. \ref{sec:Point Particle Action}, we expand the non-charged point particle action up to 2PN order and derive the Feynman rules for NRG fields coupling to worldlines. In Sec. \ref{sec:Charged Particle Action}, we do the same with charges included. In Sec. \ref{sec:Einstein-Hilbert}, we present the Einstein-Hilbert action up to 2PN order with a gauge-fixing term and derive its propagators and self-gravitational vertices. In Sec. \ref{sec:Maxwell}, we again introduce charges. In Sec. \ref{subsection:powercounting}, we explain the terms that will contribute to 2PN order and identify the relevant Feynman diagrams. Next, in Sec. \ref{section:coulomb}, we compute the Coulomb potential using the Feynman rules we have obtained. In Sec. \ref{sec:1PN}, we calculate the relevant 1PN order Feynman diagrams and summarize the Lagrangian. In Sec. \ref{sec:2PN}, we calculate the 2PN order Feynman diagrams and summarize the Lagrangian. In Sec. \ref{sec:results}, we present the Lagrangian up to 2PN order and discuss various aspects of it. A summary and conclusions are given in Sec. \ref{sec:conclusions}, along with future directions. Appendix~\ref{sec:appendix} contains a short compilation of useful formulas. 

Throughout this paper, we write $\int_{\bm k} = \int \frac{d^3 {\bm k}}{(2\pi)^3}$. The speed of light $c$ is set to 1, except in cases where we explicitly indicate the post-Newtonian counting in powers of $1/c^{2}$. Greek letters $\mu, \nu, \ldots$ denote four-dimensional spacetime coordinate indices, while lowercase Latin letters $i, j, k, \ldots$ are spatial coordinate indices. Our metric signature is $\eta_{ab} = \text{diag}(-1, +1, +1, +1)$. Our convention for the Riemann tensor is
\begin{equation}
R^{\mu}{}_{\nu\alpha\beta} = \Gamma^{\mu}{}_{\nu \beta , \alpha}
        - \Gamma^{\mu}{}_{\nu \alpha , \beta}
        + \Gamma^{\rho}{}_{\nu \beta} \Gamma^{\mu}{}_{\rho \alpha}
        - \Gamma^{\rho}{}_{\nu \alpha} \Gamma^{\mu}{}_{\rho \beta} ,
\end{equation}
where $\Gamma^{\mu}{}_{\nu \beta}$ are the Christoffel symbols, and a comma denotes a partial derivative.

The subscript $l={1,2}$ labels the binary components’ position ${\bm x}_l$, velocity ${\bm v}_l$, acceleration ${\bm a}_l$, mass $m_l$, and charge $q_l$. The separation between the binary objects is denoted by the vector ${\bm r} = {\bm x}_1 - {\bm x}_2$ with magnitude $r = |{\bm x}_1 - {\bm x}_2|$, and the unit vector ${\bm n} = {\bm r}/r$.
\section{EINSTEIN-MAXWELL THEORY}
\label{sec:Einstein-Maxwell Action}

We work within NRGR and assume the standard inspiral scale hierarchy \(r_c \ll r \ll \lambda\) (compact-object size \(r_c\), orbital separation \(r\), and GW/EM wavelength \(\lambda\)), which underpins the near–/far–zone split. In NRGR, the binary constituents are modeled as point-particle worldlines with EFT couplings, while the gravitational (and here electromagnetic) fields are dynamical degrees of freedom. The method employs PN power counting in \(v/c\) and a near–/far–zone separation: \emph{potential} (near-zone) modes carry spatial momentum \(|\mathbf{k}|\!\sim\!1/r\) with small frequency \(k^0\!\sim\!v/r\) and mediate instantaneous interactions, whereas \emph{radiation} (far-zone) modes have \(k^0\!\sim\!|\mathbf{k}|\!\sim\!v/r\) and describe waves. Finite-size effects enter via higher-dimensional worldline operators (Wilson coefficients); here we restrict to minimally coupled point charges and compute the conservative dynamics through 2PN.


In this section, we briefly review the action for Einstein-Maxwell theory for inspiraling binary black holes with charge. We work within a non-relativistic form of the Kaluza-Klein ansatz for the metric, which includes a set of non-relativistic gravitational (NRG) fields. We express the point particle action in terms of this metric and expand it up to second post-Newtonian (2PN) order, extracting the worldline couplings of the NRG fields. Similarly, we extract the worldline couplings of electromagnetic (EM) fields using the charged particle action up to 2PN order. To determine how the NRG fields interact with the worldline vertex, we derive the propagators for the NRG fields using the Einstein-Hilbert action with harmonic gauge up to 2PN order. We write the action of the EM fields in the Feynman gauge and derive their propagator. Additionally, we derive Feynman rules for the interaction of three EM fields, known as the 3-point vertex, and for the interaction of four EM fields, known as the 4-point vertex, at 2PN order. Finally, we discuss the post-Newtonian order counting rules for Feynman diagrams.

The inspiral dynamics of binary black holes with charges are described by Einstein-Maxwell theory, which has the following action:
\begin{equation}
    \label{eq:Spp}
    \mathcal{S} =  \mathcal{S}_{pp} + \mathcal{S}_q + \mathcal{S}_g + \mathcal{S}_{em},
\end{equation}
where $\mathcal{S}_{pp}$ is the point particle action, discussed in the following subsection. $\mathcal{S}_q$ is the charged particle action,
\begin{align}
	\mathcal{S}_{q} = \int d t qv^{\mu} A_{\mu} = -\int d t q A_{0} + \int dt qv^{i} A_{i},
	\label{eq:chargedpp}
\end{align}
where $q$ is the charge, $v^{\mu}$  the four-velocity, and $A^{\mu}$ the four-vector potential.
$\mathcal{S}_g$ is the Einstein-Hilbert action with a harmonic gauge term:
\begin{equation}
\label{eq:sg}
    \mathcal{S}_{g} =  \frac{1}{16 \pi} \int d^4 x \sqrt{-g} R -  \frac{1}{32 \pi G} \int d^4x\sqrt{-g}g_{\mu \nu} \Gamma^{\mu} \Gamma^{\nu},
\end{equation}
\noindent where $\Gamma^{\mu} \equiv \Gamma^{\mu}_{\rho \sigma} g^{\rho \sigma}$. $\mathcal{S}_{em}$ is the  electromagnetic action in curved spacetime with Feynman gauge,
\begin{flalign}
    \label{eq:sqaction}
	\mathcal{S}_{em} = \frac{-1}{16\pi} \int d^4 x \sqrt{-g} \left(  F^2  + 2(\partial_{\mu} A^\mu)^2 \right),
\end{flalign}
where $F_{\mu\nu} = \nabla_{\mu} A_{\nu} - \nabla_{\nu}A_{\mu}$.

We use the metric in a non-relativistic form according to the Kaluza-Klein (KK) ansatz,
\begin{equation} \label{eq:kka}
d\tau^2 = g_{\mu\nu}dx^{\mu}dx^{\nu} \equiv -e^{2 \phi}(dt - \mathcal{A}_i\, dx^i)^2 + e^{-2 \phi} \gamma_{ij}dx^i dx^j~,
\end{equation}
where ${\phi, \mathcal{A}_i,\gamma_{ij}\equiv\delta_{ij}+\sigma_{ij}}$ are a set of NRG fields.  
We can write the matrix form of the metric in terms of the NRG fields as
\begin{equation}
    g_{\mu \nu} = 
    \begin{pmatrix}
    -e^{2 \phi} & e^{2\phi} \mathcal{A}_j \\ 
    e^{2\phi} \mathcal{A}_{i} \; \; \; & e^{-2 \phi} \gamma_{ij} - e^{2 \phi} \mathcal{A}_{i} \mathcal{A}_{j} 
    \end{pmatrix}.
    \label{eq:metricKK}
\end{equation} 
There is a physical meaning associated with the NRG fields: $\phi$ can be identified with the Newtonian potential, $\mathcal{A}_{i}$ is the gravito-magnetic potential, and $\sigma_{ij}$ is the 3-metric tensor ~\cite{Kol:2010si, Kol:2007bc}. The leading PN order of NRG fields $\phi$, $\mathcal{A}_i$ and $\sigma_{ij}$ are $\mathcal{O}(1/c^2)$, $\mathcal{O}(1/c^3)$, and $\mathcal{O}(1/c^4)$ respectively.

\subsection{Point Particle Action}
\label{sec:Point Particle Action}
In the EFT approach, binary black holes are modeled as point particles moving along worldlines, governed by the action:
\begin{equation}
    \label{eq:Spp}
    \mathcal{S}_{pp} = -\sum_{i=1}^2 m_i\int d \tau_i \;= -\sum_{i=1}^2\int dt \; m_i\sqrt{-g_{\mu\nu}v_i^{\mu}v_i^{\nu}},
\end{equation}
where $m_i$, $i = 1,2$ represent the individual point particle masses. It is convenient to parameterize the worldline using coordinate time $t$. The point particle action in Eq.\eqref{eq:Spp} includes contributions from both point particles, differing only by the labels $i=1, 2$. We consider the action without this label and expand it using the KK metric from Eq.\eqref{eq:kka} up to the 2PN order,
\begin{widetext}
\begin{flalign}
\begin{split}
    \mathcal{S}_{pp} = -m \int dt &\left( 1 - \frac{1}{2}{\bm v^2} + \phi  - \mathcal{A}_iv^i - \frac{1}{8}v^4 + \frac{3}{2}\phi v^2 + \frac{1}{2}\phi^2   - \frac{1}{2} v^i v^j \sigma_{ij}   -\frac{1}{16}v^6 \right),
\end{split}
\label{eq:sppkk}
\end{flalign}
\end{widetext}
where we have used the leading PN order of the NRG fields to determine the 2PN order terms. We then extract the vertex for the NRG fields coupling to the worldline mass up to the 2PN order:  
\begin{align}
    \label{gravitonscalarvertex}
        \begin{gathered}
   \begin{tikzpicture}
\draw[black, ultra thick] (0,-1.5) -- (0,1.5);
\draw[black,  thick] (0,0) -- (1.5,0);
\filldraw[black] (0,0) circle (3pt) node[anchor=west]{};
\end{tikzpicture}
        \end{gathered}  &= - m \int d t \phi \left\{ 1 + \frac{3}{2}v^2  \right\},\\
    \label{gravitonvectorvertex}    
        \begin{gathered}
        \begin{tikzpicture}
        \draw[black, ultra thick] (0,-1.5) -- (0,1.5);
        \draw[snake=coil,segment aspect=0, thick] (0,0) -- (1.5,0);
        \filldraw[black] (0,0) circle (3pt) node[anchor=west]{};
        \end{tikzpicture}   
        \end{gathered} &= ~m \int dt~ \mathcal{A}_iv^i,\\
        \begin{gathered}
        \begin{tikzpicture}
        \draw[black, ultra thick] (0,-1.5) -- (0,1.5);
        \draw[snake=coil,segment aspect=0, thick] (0,0) -- (1.5,0);
        \draw[snake=coil,segment aspect=0, thick] (0,0.1) -- (1.5,0.1);
        \filldraw[black] (0,0) circle (3pt) node[anchor=west]{};
        \end{tikzpicture}   
        \end{gathered}&= ~\frac{m}{2} \int dt~\sigma_{ij}v^iv^j~
\end{align}
where the heavy solid lines represent the worldlines, and the black blobs represent the particle mass on the worldline. Similarly, we extract the exchange of the two scalar NRG fields:
\begin{align}
    \begin{gathered}
   \begin{tikzpicture}
\draw[black, ultra thick] (0,-1.5) -- (0,1.5);
\draw[black,  thick] (0,0) -- (1,1);
\draw[black,  thick] (0,0) -- (1,-1);
\filldraw[black] (0,0) circle (3pt) node[anchor=west]{};
\end{tikzpicture}
    \end{gathered}
    ={}& - \frac{1}{2} m \int dt  \phi^2 .
\end{align}
These are the only NRG fields that couple with the worldline particle up to 2PN order. Using a similar procedure, one can extend this to higher PN order couplings of NRG fields.
\subsection{Charged Particle Action}
\label{sec:Charged Particle Action}
The contribution to the worldline action arising from the charge is
\begin{align}
	\mathcal{S}_{q} = \int d t qv^{\mu} A_{\mu} = \int d t q A_{0} + \int dt qv^{i} A_{i}.
	\label{eq:chargedpp}
\end{align}
We find the scalar electromagnetic field worldline vertex to be
\begin{align}
    \label{emscalar}
    \begin{gathered}
   \begin{tikzpicture}
\draw[black, ultra thick] (0,-1.5) -- (0,1.5);
\draw[dashed, thick] (0,0) -- (1.5,0);
\filldraw[black] (0,0) circle (3pt) node[anchor=west]{};
\end{tikzpicture}
    \end{gathered}
    ={}& q \int dt  A_0 ,
\end{align}
\noindent Similarly, the vector electromagnetic field worldline vertex is given by:
\begin{align}
    \label{emvector}
    \begin{gathered}
   \begin{tikzpicture}
\draw[black, ultra thick] (0,-1.5) -- (0,1.5);
\draw[dashed, thick] (0,0.05) -- (1.5,0.05);
\draw[dashed, thick] (0,-0.05) -- (1.5,-0.05);
\filldraw[black] (0,0) circle (3pt) node[anchor=west]{};
\end{tikzpicture}
    \end{gathered}
    ={}& q \int dt  A_i v^i .
\end{align}
\subsection{Einstein-Hilbert Action}
\label{sec:Einstein-Hilbert}
We consider the gravitational action to consist of the Einstein-Hilbert action along with a harmonic gauge fixing term:
\begin{equation}
\label{eq:sg}
    \mathcal{S}_{g} = \frac{1}{16 \pi G} \int d^4 x \sqrt{-g} R -  \frac{1}{32 \pi G} \int d^4x\sqrt{-g}g_{\mu \nu} \Gamma^{\mu} \Gamma^{\nu},
\end{equation}
Using the KK metric from Eq.~\eqref{eq:kka}, the gravitational action $\mathcal{S}_g$ can be expressed as \cite{Kol:2007bc}:
\begin{flalign}
\mathcal{S}_g = \frac{1}{16 \pi G} \int d t d^3  x \sqrt{\gamma} \left( R[\gamma] - 2 \partial_{i} \phi \partial_{i} \phi + \frac{1}{4} e^{4 \phi} \mathcal{F}^2 \right),
\label{eq:KKgravity}
\end{flalign}
\noindent with $\partial_{i} \partial_i \phi = \gamma^{ij} \partial_i \phi \partial_j \phi$, and defining the analog of the field strength tensor $\mathcal{F}_{ij} = \partial_i \mathcal{A}_j - \partial_j \mathcal{A}_i$ using the KK gravito-magnetic tensor $\mathcal{A}_i$.\\
From the gravitational action in Eq.~\eqref{eq:KKgravity}, we derive the NRG field propagators \cite{Levi:2008nh}:
\begin{align}
    \label{scalargraviton}
    \left< \phi ({\bm x_1}, t_1) \phi ({\bm x_2},t_2)\right> &= 4 \pi G \; \delta(t_1 - t_2)  \int_{\bm k}  \frac{e^{i {\bm k}\cdot {\bm r}}}{{\bm k^2}}, \\
    \left< \mathcal{A}_i ({\bm x_1},t_1) \mathcal{A}_j({\bm x_2},t_2)\right> &= - 16\pi G \; \delta (t_1 - t_2)  \int_{\bm k} \frac{e^{i {\bm k}\cdot {\bm r}}}{{\bm k^2}} \delta_{ij}, \\
    \langle{\sigma_{ij}({\bm x_1},t_1)}{\sigma_{kl}({\bm x_2},t_2)}\rangle &= 32\pi G~P_{ij;kl} ~\delta(t_1-t_2)\int_{\bm k} \frac{e^{i {\bm k}\cdot {\bm r}}}{\bm k^2},
\end{align}
where $P_{ij;kl} \equiv \frac{1}{2} \left(\delta_{ik}\delta_{jl} + \delta_{il}\delta_{jk} - 2\delta_{ij}\delta_{kl}\right)$, ${\bm r} = {\bm x}_1 - {\bm x}_2$, and the propagator is instantaneous \cite{Levi:2018nxp},
\begin{equation}
    \int \frac{dk_0}{2\pi}e^{-ik_0t}\int \frac{d^3{\bm k}}{(2\pi)^3}\frac{e^{i {\bm k}\cdot{\bm x}}}{{\bm k^2}}=\delta(t)\int_{\bm k}\frac{e^{i {\bm k}\cdot{\bm x}}}{{\bm k^2}}
\end{equation}
where we use the scaling of the orbital modes, $k_0 \sim v/r$ and $|{\bm k}| \sim 1/r$.
Additionally, we obtain corrections to the instantaneous nature of the non-relativistic potential propagators. The Feynman rules for these propagator correction vertices are given by:
\begin{align}
    & \begin{gathered}
     \begin{tikzpicture}
    \draw[black,  thick] (0,0) -- (2,0);
    \node [cross out,draw=black, thick] at (1.,0){};
    \end{tikzpicture}
    \end{gathered} = \frac{1}{2!}\frac{1}{4\pi G} \int d^4 x ~ \partial^{0}\phi\partial^{0}\phi
\end{align}
Here, the crosses denote self-gravitational quadratic vertices, containing two time derivatives. Typically, we omit the factors $1/i!$ for $i$ identical lines on a vertex, instead multiplying the diagram by 1/(symmetry factor). From now on, we present the vertex rules without the $1/i!$ factors:
\begin{align}
    & \begin{gathered}
     \begin{tikzpicture}
    \draw[black,  thick] (0,0) -- (2,0);
    \node [cross out,draw=black, thick] at (1.,0){};
    \end{tikzpicture}
    \end{gathered} = \frac{1}{4\pi G} \int d^4 x ~ \partial^{0}\phi\partial^{0}\phi
\end{align}
\subsection{Electromagnetic action}
\label{sec:Maxwell}
We consider the electromagnetic action in curved spacetime with a Feynman gauge fixing term:
\begin{flalign}
    \label{eq:sqaction}
	\mathcal{S}_{em} = \frac{-1}{16\pi} \int d^4 x \sqrt{-g} \left(  F^2  + 2(\partial_{\mu} A^\mu)^2 \right),
\end{flalign}
where $F_{\mu\nu} = \nabla_{\mu} A_{\nu} - \nabla_{\nu} A_{\mu}=\partial_{\mu} A_{\nu} - \partial_{\nu} A_{\mu}$. 
Since, the covariant derivative expansion into partial derivatives and Christoffel symbols results in the cancellation of the Christoffel symbols, leaving only partial derivatives.
We further expand the action in Eq.\eqref{eq:sqaction} using the KK metric from Eq.\eqref{eq:kka} up to 2PN order:
\begin{widetext}
\begin{align}
\label{eq:maxwellmetric}
\mathcal{S}_q &= \frac{-1}{16\pi} \int d^4 x \Bigg[ -2 (\partial^{i}A_{0}) (\partial_{i}A_{0}) + 2 (\partial_0 A_0)^2 + 2 \delta^{lk} (\partial^i A_k)(\partial_i A_l) + 4 \phi (\partial^{i}A_{0})(\partial_{i}A_{0}) - 2 \delta^{ij} (\partial_{0}A_{j})(\partial_{0}A_{i}) - 12 \phi (\partial_0 A_0)^2 \nonumber \\
& \quad + 4 \phi \delta^{lk} (\partial^i A_k)(\partial_i A_l) + 2 \sigma^{ij} (\partial_{j}A_{0})(\partial_{i}A_{0}) - \sigma_i^i (\partial^{i}A_{0})(\partial_{i}A_{0}) + 4 \mathcal{A}^j (\partial_{k}A_{j})(\partial^{k}A_{0}) + 4 (\partial^i \mathcal{A}^j) (\partial_j A_i) A_0 \nonumber \\
& \quad - 4 (\partial_j \mathcal{A}^j) (\partial^i A_i) A_0 + 8 (\partial^j \phi)\partial^i A_j A_i - 8\partial^i \phi(\partial^j A_j A_i) - 4 \mathcal{A}^j (\partial_0 A_0)(\partial_j A_0) - 4 \phi^2 (\partial^{i}A_{0})(\partial_{i}A_{0}) \Bigg]
\end{align}
\end{widetext}
This action agrees with the one resulting from a calculation with the EFTofPNG package \cite{Levi:2017kzq}. We find the electromagnetic field propagators from the first and second terms of Eq.\eqref{eq:maxwellmetric} to be
\begin{align}
    \label{emscalarprop}
    \left< A_0 ({\bm x_1},t_1) A_0 ({\bm x_2},t_2)\right> = -4\pi  \; \delta(t_1 - t_2)  \int_{\bm k} \frac{e^{i {\bm k}\cdot {\bm r}}}{{\bm k^2}} ,\\
    \label{emvectorprop}
    \left< A_i ({\bm x_1},t_1) A_j({\bm x_2},t_2)\right> = 4 \pi \; \delta (t_1 - t_2)  \int_{\bm k} \frac{e^{i {\bm k}\cdot {\bm r}}}{{\bm k^2}} \delta_{ij},
\end{align}    
yielding corrections to the instantaneous nature of the non-relativistic potential propagators. The Feynman rules for the propagator correction vertices come from the third and fifth terms of Eq.\eqref{eq:maxwellmetric},
\begin{align}
    \label{timederivative}
    & \begin{gathered}
     \begin{tikzpicture}
\draw[dashed, thick] (0,0) -- (2,0);
\node [cross out,draw=black, thick] at (1.,0){};
\end{tikzpicture}
    \end{gathered} =  \frac{-1}{4\pi} \int d^4 x ~ \partial_0 A_0\partial_0 A_0,\\
    & \begin{gathered}
     \begin{tikzpicture}
\draw[dashed, thick] (0,0) -- (2,0);
\draw[dashed, thick] (0,0.1) -- (2,0.1);
\node [cross out,draw=black, thick] at (1.,0.05){};
\end{tikzpicture}
    \end{gathered} =   \frac{1}{4\pi} \int d^4 x ~ \partial_{0}A_{i}\partial_{0}A_{j}\delta^{ij},
\end{align}
where the crosses represent the self-electromagnetic quadratic vertices, containing two time derivatives.\\
We extract the three-point self-interacting vertex from Eq.~\eqref{eq:maxwellmetric},
\begin{widetext}
\begin{align}
\label{fig:3pointv1}
    & \begin{gathered}
     \begin{tikzpicture}
\draw[black,  thick] (0,0) -- (1.,1.);
\draw[dashed,  thick] (0,0) -- (1.,-1.);
\draw[dashed,  thick] (-1,0) -- (0,0);
\filldraw[black] (0,0) circle (3pt) node[anchor=west]{};
\end{tikzpicture}
    \end{gathered} =  \frac{-1}{8\pi} \int d^4x \; \left(4\phi(\partial_{i}A_{0})(\partial^{i}A_{0})-12\phi\partial_0 A_0\partial_0 A_0\right),\\
    \label{fig:3pointv2}
    & \begin{gathered}
     \begin{tikzpicture}
\draw[dashed,  thick] (0,0) -- (1.,1.);
\draw[dashed,  thick] (0,-0.1) -- (1.1,1.);
\draw[dashed,  thick] (0,0) -- (1.,-1.);
\draw[dashed,  thick] (-0.1,0) -- (1.,-1.1);
\draw[black,  thick] (-1,0) -- (0,0);
\filldraw[black] (0,0) circle (3pt) node[anchor=west]{};
\end{tikzpicture}
    \end{gathered} =  \frac{-1}{8\pi}  \int d^4x \; 4 \phi\delta^{lk} \partial^i A_k\partial_i A_l+8( \partial^j \phi \partial^i A_j A_i -  \partial^i \phi \partial^j A_j A_i),\\
    \label{fig:3pointv3}
    & \begin{gathered}
     \begin{tikzpicture}
\draw[dashed,  thick] (0,0) -- (1.,1.);
\draw[dashed,  thick] (0,-0.1) -- (1.1,1.);
\draw[snake=coil,segment aspect=0,  thick] (0,0) -- (1.,-1.);
\draw[dashed,  thick] (-1,0) -- (0,0);
\filldraw[black] (0,0) circle (3pt) node[anchor=west]{};
\end{tikzpicture}
    \end{gathered} =  \frac{-1}{16\pi}  \int d^4x \; 4 \mathcal{A}^j (\partial_{k}A_{j})(\partial^{k}A_{0})+ 4\partial^i \mathcal{A}^j \partial_j A_i A_0 - 4\partial_j \mathcal{A}^j \partial^i A_i A_0 ,\\
    \label{fig:3pointv4}
    & \begin{gathered}
     \begin{tikzpicture}
\draw[dashed,  thick] (0,0) -- (1.,1.);
\draw[snake=coil,segment aspect=0,  thick] (0,0) -- (1.,-1.);
\draw[dashed,  thick] (-1,0) -- (0,0);
\filldraw[black] (0,0) circle (3pt) node[anchor=west]{};
\end{tikzpicture}
    \end{gathered} =  \frac{-1}{8\pi}  \int d^4x \; (-4) \mathcal{A}_j\partial_0 A_0\partial_j A_0,\\
    \label{fig:3pointv5}
    & \begin{gathered}
     \begin{tikzpicture}
\draw[snake=coil,segment aspect=0,  thick] (0,0) -- (1.,1.);
\draw[snake=coil,segment aspect=0,  thick] (0,-0.1) -- (1.1,1.);
\draw[dashed,  thick] (0,0) -- (1.,-1.);
\draw[dashed,  thick] (-1,0) -- (0,0);
\filldraw[black] (0,0) circle (3pt) node[anchor=west]{};
\end{tikzpicture}
    \end{gathered} =   \frac{-1}{8\pi} \int d^4x \; 2 \sigma_{ij} (\partial_{j}A_{0})(\partial_{i}A_{0})-(\partial_{i}A_{0})^2\sigma_i^i.
\end{align}
\end{widetext}
We extract a four-point vertex from the last term of Eq.~\eqref{eq:maxwellmetric},
\begin{align}
    \label{4pointvertex}
    & \begin{gathered}
     \begin{tikzpicture}
\draw[black,  thick] (0,0) -- (1.,1.);
\draw[dashed,  thick] (-1.,1.) -- (0,0);
\draw[black,  thick] (0,0) -- (1.,-1.);
\draw[dashed,  thick] (-1.,-1.) -- (0,0);
\filldraw[black] (0,0) circle (3pt) node[anchor=west]{};
\end{tikzpicture}
    \end{gathered} =   \frac{-1}{4\pi} \int d^4x \; (-4)(\partial_{i}A_{0})(\partial^{i}A_{0}) \phi ^2.
\end{align}
There are two kinds of two identical lines on the vertex, and we have dropped the factors $1/2!$ and $1/2!$. 

\subsection{Power counting and Feynman diagrams}
\label{subsection:powercounting}
For a bound state, the virial theorem relates the orbital velocity $v$, Newton’s constant $G$, and charges $q$ through the relationship $v^2 \sim Gm/r \sim q^2/r$. Using this relationship, we count the powers of $G$, $q^2$, and $v^2$ separately to determine the corresponding PN order. Table \ref{fig:table1} shows that the diagram responsible for the Coulomb potential scales as $\mathcal{O}(q^2)$. We also calculate 1PN order diagrams, which scale as $\mathcal{O}(q^2v^2)$ and $\mathcal{O}(Gq^2)$. For the 2PN potential calculation, we include all diagrams that scale as $\mathcal{O}(q^2v^4)$, $\mathcal{O}(Gq^2v^2)$, $\mathcal{O}(G^2q^2)$, and $\mathcal{O}(Gq^4)$. Diagrams without charge interaction have already been computed; see \cite{Gilmore:2008gq}.

\begin{table}
    \centering
    \resizebox{0.4\textwidth}{!}{%
    \begin{tikzpicture}
      \draw[red, thick] (-3,-2.5) -- (-3,2.5);
      \draw[red, thick] (-1,-2.5) -- (-1,2.5);
      \draw[red, thick] (1,-2.5) -- (1,2.5);
      \draw[red, thick] (3,-2.5) -- (3,2.5);
      \draw[red, thick] (5,-2.5) -- (5,2.5);
      \draw[red, thick] (-3,2.5) -- (5,2.5);
      \draw[red, thick] (-3,-2.5) -- (5,-2.5);
      \filldraw[red] (-3,1.5) rectangle (5,1.5);
      \filldraw[red, fill opacity=0.3] (-3,1.5) rectangle (5,2.5);
      \node[scale=1.5, color=blue] at (-2,2) {0PN};
      \node[scale=1.5] at (-2,1) {$q^2$};
      \node[scale=1.5, color=blue] at (0,2) {1PN};
      \node[scale=1.5] at (0,1) {$q^2v^2$};
      \node[scale=1.5] at (0,0) {$Gq^2$};
      \node[scale=1.5, color=blue] at (2,2) {2PN};
      \node[scale=1.5] at (2,1) {$q^2v^4$};
      \node[scale=1.5] at (2,0) {$Gq^2v^2$};
      \node[scale=1.5] at (2,-1) {$G^2q^2$};
      \node[scale=1.5] at (2,-2) {$Gq^4$};
      \node[scale=1.5, color=blue] at (4,2) {3PN};
      \node[scale=1.5] at (4,1) {$q^2v^6$};
      \node[scale=1.5] at (4,0) {$Gq^2v^4$};
      \node[scale=1.5] at (4,-1) {$G^2q^2v^2$};
      \node[scale=1.5] at (4,-2) {$Gq^4v^2$};
    \end{tikzpicture}
    }
    \caption{This table shows the order of the terms that contribute at given PN orders.}
    \label{fig:table1}
\end{table}
We begin by generating all the relevant Feynman diagram topologies up to $G^3$, following the rules for counting powers of $G$. Specifically, when $n$ gravitons are attached to a worldline, the corresponding power is $G^{n/2}$, and each $n$-graviton self-interaction vertex carries a power of $G^{(n/2-1)}$ \cite{Gilmore:2008gq}. After determining the powers of $G$, we proceed to count the powers of velocity for each Feynman diagram.

The rules for counting powers of $v^2$ for NRG fields $\phi$, $\mathcal{A}_i$, and $\sigma_{ij}$ coupling to the worldlines are $\mathcal{O}(v^0)$, $\mathcal{O}(v^1)$, and $\mathcal{O}(v^2)$, respectively. For the EM fields, the couplings of $A_0$ and $A_i$ are $\mathcal{O}(v^0)$ and $\mathcal{O}(v^1)$, respectively. Additionally, there can be couplings of 2 or 3 EM fields to the worldlines with various orders in velocity, as discussed in Sec.~\ref{sec:Charged Particle Action}. Time derivatives can arise from either purely gravitational interaction vertices or electromagnetic interaction vertices, where each time derivative insertion is counted as $\mathcal{O}(v^2)$. Finally, we count the contributions of $q^2$; all these contributions stem from the fields coupling to the worldlines as detailed in Sec.~\ref{sec:Charged Particle Action}.

We generate all relevant diagram topologies and populate them with the three NRG fields $\phi$, $\mathcal{A}_i$, and $\sigma_{ij}$, as well as the EM fields $A_0$ and $A_i$ in all allowed combinations. Next, we count the factors of $v$ resulting from the gravitational and electromagnetic fields coupling to the worldline, and any time derivatives acting on internal vertices. We also count the factors of $q$ arising from fields coupling with worldlines. After determining the powers of $G$, $v^2$, and $q^2$, we identify the relevant diagrams at a given PN order. The diagrams relevant at 0PN order are shown in Fig.\ref{fig:coulombfig}, those at 1PN order in Fig.\ref{fig:fig1}, and those at 2PN order in Figs.~\ref{fig:q^2v^4}, \ref{fig:Gq^2v^2}, \ref{fig:Gq^4}, and \ref{fig:G^2q^2}. We ignore diagrams with gravitational loops since they include quantum effects \cite{Holstein:2004dn}.

\section{Coulomb Potential}
\label{section:coulomb}

In this section, we compute the Coulomb potential term using the relevant Feynman diagram.
\begin{figure}[hbt!]
\centering
\includegraphics{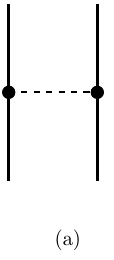} 
\caption{Diagram (a) shows the Coulomb interaction at $\mathcal{O}(q^2)$}
\label{fig:coulombfig}  
\end{figure}
We start with diagram 1(a) in Fig.\ref{fig:coulombfig}(a). To compute this diagram, we need two copies of the worldline coupling with the EM scalar field in Eq.\eqref{emscalar} and the propagator for the scalar EM field in Eq.~\eqref{emscalarprop}:
\begin{align}
    \mathrm{Diagram} \; \ref{fig:coulombfig} \mathrm{(a)}
    = & ~q_1q_2 \int dt_1\int dt_2  \left<A_0({\bm {\bm x_1}},t_1) A_0({\bm x_2},t_2)\right> \nonumber &\\
    = & q_1q_2 \int dt_1\int dt_2  (-4\pi)  \; \delta(t_1 - t_2)  \int_{\bm k} \frac{e^{i {\bm k}\cdot {\bm r}}}{{\bm k^2}} \nonumber &\\
    = & -4\pi q_1q_2  \int dt \frac{1}{4\pi r} = \int dt \frac{-q_1q_2}{r}.
\end{align}
We obtain the Coulomb potential term after performing the Fourier integral using identity (\ref{eq:ft}).
\section{1 PN order Potential}
\label{sec:1PN}
\begin{figure}[hbt!]
\centering
\includegraphics{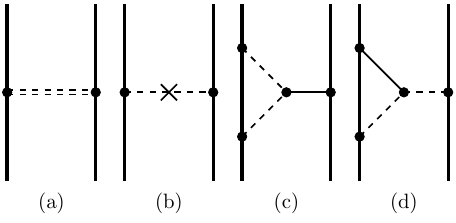} 
\caption{Diagrams (b) and (c) are 1PN corrections at $\mathcal{O}(q^2v^2)$, and the rest of the diagrams are also 1PN corrections at $\mathcal{O}(Gq^2)$} 
\label{fig:fig1}  
\end{figure}

In this section, we compute the 1PN Feynman diagrams, which include $\mathcal{O}(Gq^2)$ and $\mathcal{O}(v^2q^2)$ contributions. We explicitly demonstrate the computation of diagrams 2(b) in Fig.\ref{fig:fig1}, which include a time derivative, and diagrams 2(c) in Fig.\ref{fig:fig1}, which involve a loop integral. The symmetry factors of the diagrams are determined as described in \cite{Gilmore:2008gq}; alternatively, they can be computed using Wick’s theorem without considering time ordering. We specify the symmetry factor for a diagram when it is not equal to 1. Finally, the 1PN Lagrangian is obtained by summing the contributions from all the 1PN diagrams.
\subsection{$q^2v^2$ order diagrams}
Diagram 2(a) in Fig.\ref{fig:fig1} is computed using two copies of worldline couplings with the vector EM field in Eq.\eqref{emvector}, the propagator for the vector EM field in Eq.\eqref{emvectorprop}, and the formula in Eq.\ref{eq:ft}:
\begin{flalign}
\begin{split}
\mathrm{Diagram} &\; \ref{fig:fig1} \mathrm{(a)} = \int dt_1 \;q_1 v^i_1\int dt_2 \;q_2 v^j_2\left< A_i ({\bm x_1},t_1) A_j ({\bm x_2},t_2)\right> \\
=& \;q_1 q_2 \int dt_1 \; v^i_1\int dt_2 \; v^j_2 \;4 \pi \; \delta (t_1 - t_2)  \int_{\bm k} \frac{e^{i {\bm k}\cdot {\bm r}}}{{\bm k^2}} \delta_{ij} \\
=& \int dt \frac{q_1 q_2}{ r} \left({\bm v}_1 \cdot {\bm v}_2 \right). \\
\end{split}
\end{flalign}
Diagram 2(b) in Fig.\ref{fig:fig1} contains a time derivative. It is computed using Eq.\eqref{emscalar} for the worldline couplings, Eq.\eqref{timederivative} for time derivative terms, and Eq. \eqref{emscalarprop} for the scalar EM field propagator. We obtain the following:
\begin{widetext}
\begin{align}
\begin{split}
	\mathrm{Diagram \; } \ref{fig:fig1} \mathrm{(b)} &=  -4\pi q_1 q_2 \int d^4 x \int dt_1 \int dt_2 ~ \partial_t \delta(t_1 - t) \partial_t \delta(t_1 - t) \int_{\bm k_1} \frac{e^{i{\bm k_1} \cdot ({\bm x_1}-{\bm x})}}{\bm k_1^2} \int_{\bm k_2} \frac{e^{i{\bm k_2} \cdot ({\bm x_2}-{\bm x})}}{{\bm k_2^2}}.
\end{split}
\end{align}
\end{widetext}

Performing the integration over $d^3 x$ results in a delta function in momentum. Next, the momentum integral involving this delta function is evaluated, and the time derivative is flipped using identity (\ref{eq:timeflip}). Finally, the integration is completed using the tensor Fourier identity (\ref{tensorfourierindentity}), which gives:

\begin{align}
    \mathrm{Diagram \; } \ref{fig:fig1} \mathrm{(b)} &=   - \int d t \frac{q_1 q_2}{2r} \left( {\bm v}_1 \cdot {\bm v}_2 - ({\bm v}_1 \cdot {\bm n})({\bm v}_2 \cdot {\bm n}) \right).
\end{align}

\subsection{$Gq^2$ order diagrams}
Diagram 2(c) in Fig.\ref{fig:fig1} is complex to compute, so we will address it step by step. We start by using Eq.\eqref{emscalar} and Eq.\eqref{gravitonscalarvertex} for the worldline couplings, along with the 3-point vertex from Eq.\eqref{fig:3pointv1}. We then apply Eq. \eqref{emscalarprop} for the scalar EM field propagator and Eq.\eqref{scalargraviton} for the scalar NRG field propagators. After applying these equations, we perform the integration over $d^3 x$ and time, which results in:
\begin{flalign}
\begin{split}
	\mathrm{Diagram \; } \ref{fig:fig1} \mathrm{(c)} 
    &= \alpha  \int dt  \int_{k_2}  \int_{k_3}  \; \frac{e^{i {\bm k_3}\cdot ({{\bm x_2}}-{{\bm x_1}})}}{({\bm k_2}+{\bm k_3})^2}\; \frac{(k_2^i+k_3^i)k_2^j}{{\bm k_2}^2{\bm k_3}^2}\;\delta_{ij}.\\
\end{split}
\end{flalign}
where $\alpha = (4\pi)^2\;Gm_2q_1^2$. This integral is solved using the 1-loop tensor master integral (\ref{eq:1loopvec}) and the tensor Fourier identity (\ref{tensorfourierindentity}), which yield
\begin{flalign}
\begin{split}
	\mathrm{Diagram \; } \ref{fig:fig1} \mathrm{(c)} 
    &=     -\int dt \; \frac{Gm_2q_1^2}{r^2}.
\end{split}
\end{flalign}
The symmetry factor of this diagram is $\frac{1}{2}$, which needs to be multiplied,
\begin{equation}
    \mathrm{Diagram \; } \ref{fig:fig1} \mathrm{(c)} =     -\int dt \; \frac{Gm_2q_1^2}{2r^2}.
\end{equation}
We compute the diagram 2(d) similarly,
\begin{flalign}
\begin{split}
	\mathrm{Diagram \; } \ref{fig:fig1} \mathrm{(d)} 
    = \int dt \frac{Gm_2q_1q_2}{ r^2}.
\end{split}
\end{flalign}

Combining the contributions from all 1PN diagrams and including the corresponding contributions from interchanging $1 \leftrightarrow 2$ and $\bm{n} \rightarrow -\bm{n}$, which results in different diagrams. we obtain the  Lagrangian including the kinetic energy term
\begin{widetext}
\begin{equation}
    \mathcal{L}_{1PN} = \frac{1}{8}m_1v_1^4 + \frac{1}{8}m_2v_2^4 + \frac{q_1 q_2}{ r} \left({\bm v}_1 \cdot {\bm v}_2 \right) - \frac{q_1 q_2}{2r} \left( {\bm v}_1 \cdot {\bm v}_2 - ({\bm v}_1 \cdot {\bm n})({\bm v}_2 \cdot {\bm n}) \right)+ \frac{Gq_1q_2}{ r^2}(m_2+m_1)-\frac{G}{2r^2}(m_2q_1^2+m_1q_2^2),
\end{equation}
\end{widetext}
This 1PN Lagrangian agrees with what was obtained in \cite{Patil:2020dme, Khalil:2018aaj}. The $\mathcal{O}(G^0)$ part of this Lagrangian is known as the Darwin Lagrangian, first derived in 1920, which describes the leading relativistic corrections to the interaction between charged particles.

\section{2PN order}
\label{sec:2PN}
In this section, we compute the 2PN Feynman diagrams. At 2PN order, the relevant diagrams fall into four categories: $\mathcal{O}(q^2v^4)$, $\mathcal{O}(Gq^2v^2)$, $\mathcal{O}(Gq^4)$, and $\mathcal{O}(G^2q^2)$. These are detailed in the subsections below.

The $\mathcal{O}(q^2v^4)$ diagrams involve time derivatives and are computed similarly to diagram 2(b) in Fig.~\ref{fig:fig1}. The $\mathcal{O}(Gq^2v^2)$ diagrams include one-loop integrals and a time derivative, analogous to the 1PN order diagrams. The $\mathcal{O}(Gq^4)$ and $\mathcal{O}(G^2q^2)$ diagrams involve two-loop integrals, which can be categorized into three types.

The first type consists of factorizable two-loop integrals that decompose into a product of two one-loop integrals, allowing each one-loop to be computed separately. These integrals are relatively easier to compute. The second type includes nested two-loops, where a one-loop is nested within another loop, requiring successive computation—first the nested one-loop, then the outer one-loop. The third type comprises irreducible two-loops, which can be formally reduced, using integration by parts method \cite{Smirnov:2004ym}, to a sum of factorizable and nested two-loops.

We specify the symmetry factor for each diagram when it is not equal to 1 and present the results for diagrams with the symmetry factor included.

\begin{figure}[htbp] 
   \centering
   \includegraphics{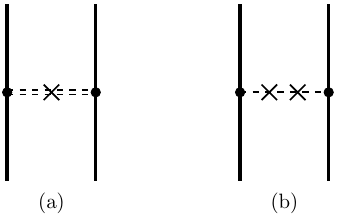} 
\caption{Diagrams that have $\mathcal{O}(q^2v^4)$ and time derivatives.}
\label{fig:q^2v^4}
\end{figure}

\subsection{$q^2v^4$ order diagrams}
There are only two diagrams of this order, neither of which involves graviton exchange, as shown in Fig.~\ref{fig:q^2v^4}. These diagrams contain time derivatives and are computed similarly to diagram 2(b):

\begin{widetext}
\begin{align}
\mathrm{Diagram \;} \ref{fig:q^2v^4} \mathrm{(a)} &=   \int d t \frac{q_1 q_2}{2r} \left[ \left( {\bm v}_1 \cdot {\bm v}_2 - ({\bm v}_1 \cdot {\bm n})({\bm v}_2 \cdot {\bm n}) \right) ({\bm v}_1 \cdot {\bm v}_2) -   {\bm a}_1 \cdot {\bm a}_2 \;r^2 - ({\bm v}_1 \cdot {\bm a}_2)  ({\bm v}_1 \cdot {\bm r}) + ({\bm v}_2 \cdot {\bm a}_1)  ({\bm v}_2 \cdot {\bm r})  \right]
\end{align}
and 
\begin{align}
    \mathrm{Diagram \; } \ref{fig:q^2v^4} \mathrm{(b)} &= -\int d t\; \frac{q_1 q_2}{8r} \left[3({\bm n} \cdot {\bm v}_1)^2({\bm n} \cdot {\bm v}_2)^2 + 2({\bm v}_1 \cdot {\bm v}_2)^2 + {\bm v}_1^2{\bm v}_2^2 - 4({\bm n}\cdot{\bm v}_1) ({\bm n}\cdot {\bm v}_2) ({\bm v}_1 \cdot {\bm v}_2) - ({\bm n}\cdot {\bm v}_1)^2{\bm v}_2^2   \right. \nonumber \\ & \left.-({\bm n}\cdot {\bm v}_2)^2{\bm v}_1^2 - ({\bm a}_1 \cdot {\bm n})({\bm a}_2 \cdot {\bm n})r^2 - ({\bm a}_1 \cdot {\bm a}_2)r^2 - 2 ({\bm n} \cdot {\bm v}_1) ({\bm a}_2 \cdot {\bm v}_1) - ({\bm a}_2 \cdot {\bm n}){\bm v}_1^2 + ({\bm v}_1 \cdot {\bm n})^2({\bm a}_2 \cdot {\bm n}) \right. \nonumber \\ & \left. +2 ({\bm n} \cdot {\bm v}_2) ({\bm a}_1 \cdot {\bm v}_2) + ({\bm a}_1 \cdot {\bm n}){\bm v}_2^2 - ({\bm v}_2 \cdot {\bm n})^2({\bm a}_1 \cdot {\bm n})\right]
\end{align}

\end{widetext}

\subsection{$Gq^2v^2$ order diagrams}
\begin{figure*} 
   \centering
   \includegraphics{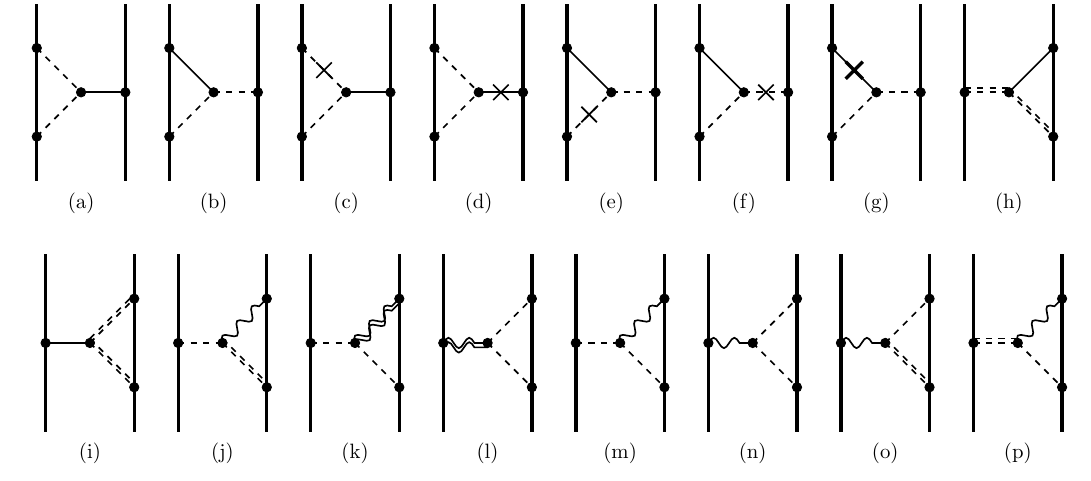} 
\caption{Diagrams that have $\mathcal{O}(Gq^2v^2)$.}
\label{fig:Gq^2v^2}
\end{figure*} 

At the $\mathcal{O}(Gq^2v^2)$ order, there are 16 diagrams to consider. Each of these diagrams features a one-loop integral with a 3-point vertex and is computed similarly to the 1PN order diagrams, involving both a time derivative and a single graviton exchange. Notably, diagrams 4(a), 4(d), 4(i), 4(l), and 4(n) in Fig.~\ref{fig:Gq^2v^2} have a symmetry factor of 1/2, while the remaining diagrams have a symmetry factor of 1. The results obtained from these diagrams are as follows:


\begin{widetext}
\begin{flalign}
\begin{split}
	\mathrm{Diagram \; } \ref{fig:Gq^2v^2} \mathrm{(a)}
	&= \frac{3}{4}Gq_1^2m_2 \int d t \frac{\left({\bm v}_1 \cdot {\bm v}_1 - {\bm v}_2 \cdot {\bm v}_2 -({\bm n} \cdot {\bm v}_1)^2\right)}{r^2},
\end{split}
\end{flalign}
\begin{flalign}
\begin{split}
	\mathrm{Diagram \; } \ref{fig:Gq^2v^2} \mathrm{(b)}
	&= -3Gq_1m_1q_2 \int d t \frac{1}{r^2}\left({\bm v}_1 \cdot {\bm v}_2-\frac{1}{2}{\bm v}_1 \cdot {\bm v}_1-2({\bm n} \cdot {\bm v}_1)({\bm n} \cdot {\bm v}_2)\right),
\end{split}
\end{flalign}

\begin{flalign}
\begin{split}
	\mathrm{Diagram \; } \ref{fig:Gq^2v^2} \mathrm{(c)}
	&= -\frac{1}{4}Gq_1^2m_2 \int d t \frac{({\bm v}_1 \cdot {\bm v}_1+({\bm n} \cdot {\bm v}_1)({\bm n} \cdot {\bm v}_1))}{r^2} ,
\end{split}
\end{flalign}


\begin{flalign}
\begin{split}
	\mathrm{Diagram \; } \ref{fig:Gq^2v^2} \mathrm{(d)}
	&= -\frac{1}{2}Gq_2^2m_1 \int d t \frac{-2({\bm n} \cdot {\bm v}_1)({\bm n} \cdot {\bm v}_2)+({\bm v}_1 \cdot {\bm v}_2)}{r^2},
\end{split}
\end{flalign}


\begin{flalign}
\begin{split}
	\mathrm{Diagram \; } \ref{fig:Gq^2v^2} \mathrm{(e)}
	&= Gq_1q_2m_1 \int d t \left[\frac{(-{\bm v}_1 \cdot {\bm v}_1+4({\bm v}_1 \cdot {\bm v}_2)}{4r^2} + \frac{(3({\bm v}_1 \cdot {\bm n})({\bm v}_1 \cdot {\bm n})-8({\bm n} \cdot {\bm v}_1)({\bm n} \cdot {\bm v}_2))}{4r^2}\right],
\end{split}
\end{flalign}
\end{widetext}
\begin{flalign}
\begin{split}
	\mathrm{Diagram \; } \ref{fig:Gq^2v^2} \mathrm{(f)}
	&= Gq_1q_2m_1 \int d t \frac{-2({\bm n} \cdot {\bm v}_1)({\bm n} \cdot {\bm v}_2)+({\bm v}_1 \cdot {\bm v}_2)}{r^2},
\end{split}
\end{flalign}


\begin{flalign}
\begin{split}
	\mathrm{Diagram \; } \ref{fig:Gq^2v^2} \mathrm{(g)}
	&= \frac{1}{4}Gq_1q_2m_1 \int d t \frac{({\bm v}_1 \cdot {\bm v}_1+({\bm n} \cdot {\bm v}_1)({\bm n} \cdot {\bm v}_1))}{r^2} 
\end{split}
\end{flalign}

\begin{flalign}
\begin{split}
	\mathrm{Diagram \; } \ref{fig:Gq^2v^2} \mathrm{(h)}
	&= Gq_1q_2m_2 \int d t \frac{({\bm v}_1 \cdot {\bm v}_2)}{r^2},
\end{split}
\end{flalign}

\begin{flalign}
\begin{split}
	\mathrm{Diagram \; } \ref{fig:Gq^2v^2} \mathrm{(i)}
	&= -\frac{1}{2}Gq_2^2m_1 \int d t \frac{({\bm v}_2 \cdot {\bm v}_2)}{r^2},
\end{split}
\end{flalign}

\begin{flalign}
\begin{split}
	\mathrm{Diagram \; } \ref{fig:Gq^2v^2} \mathrm{(j)}
	&= -2Gq_1q_2m_2 \int d t \frac{{\bm v}_2^2}{r^2},
\end{split}
\end{flalign}

\begin{flalign}
\begin{split}
	\mathrm{Diagram \; } \ref{fig:Gq^2v^2} \mathrm{(k)}
	&= 2Gq_1q_2m_2 \int d t \frac{-v^2_2+2({\bm v}_2 \cdot n) ({\bm v}_2 \cdot n)}{r^2},
\end{split}
\end{flalign}

\begin{flalign}
\begin{split}
	\mathrm{Diagram \; } \ref{fig:Gq^2v^2} \mathrm{(l)}
	&= \frac{1}{2}Gq_2^2m_1 \int d t \frac{v^2_1-({\bm v}_1 \cdot n) ({\bm v}_1 \cdot n)}{r^2},
\end{split}
\end{flalign}

\begin{widetext}
\begin{flalign}
\begin{split}
	\mathrm{Diagram \; } \ref{fig:Gq^2v^2} \mathrm{(m)}
	&= Gq_1q_2m_2 \int d t \frac{2({\bm v}_1 \cdot {\bm v}_2)+ 2({\bm v}_2 \cdot {\bm v}_2)-4({\bm v}_1 \cdot n ) ({\bm v}_2 \cdot n)- 4 ({\bm v}_2 \cdot n ) ({\bm v}_2 \cdot n)}{r^2}
\end{split}
\end{flalign}
\end{widetext}

\begin{flalign}
\begin{split}
	\mathrm{Diagram \; } \ref{fig:Gq^2v^2} \mathrm{(n)}
	&= Gq_2^2m_1 \int d t  \frac{-({\bm v}_1 \cdot {\bm v}_2)+({\bm v}_1 \cdot n) ({\bm v}_2 \cdot n) }{r^2} ,
\end{split}
\end{flalign}

\begin{flalign}
\begin{split}
	\mathrm{Diagram \; } \ref{fig:Gq^2v^2} \mathrm{(o)}
	&= 2Gq_2^2m_1 \int d t \frac{({\bm v}_1 \cdot {\bm v}_2)}{r^2},
\end{split}
\end{flalign}

\begin{flalign}
\begin{split}
	\mathrm{Diagram \; } \ref{fig:Gq^2v^2} \mathrm{(p)}
	&= -2Gq_1q_2m_1 \int d t \frac{({\bm v}_1 \cdot {\bm v}_2)}{r^2}.
\end{split}
\end{flalign}

\subsection{$Gq^4$ order diagrams}
\label{subsection:Gq^4}
\begin{figure*}
    \centering
       \includegraphics{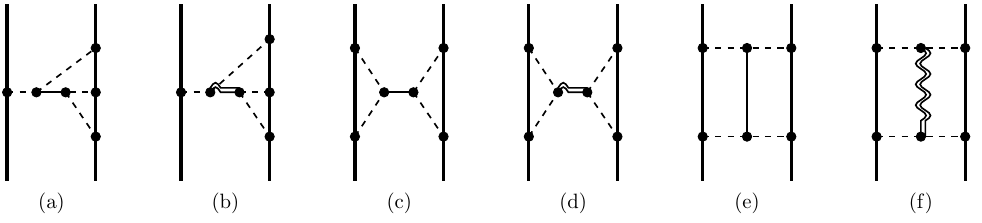} 
    \caption{Diagrams that have $\mathcal{O}(Gq^4)$.}
    \label{fig:Gq^4}
\end{figure*}
At the $\mathcal{O}(Gq^4)$ order, there are 6 two-loop diagrams. The symmetry factors for diagrams 5(a), 5(b), 5(e), and 5(f) in Fig.~\ref{fig:Gq^4} are 1/2, while those for 5(c) and 5(d) are 1/4. Diagrams 5(a) and 5(b) are nested 2-loop types, 5(c) and 5(d) are factorizable 2-loop types, and 5(e) and 5(f) are irreducible 2-loop types.


Irreducible 2-loop diagrams 5(e) and 5(f) involve two 3-point vertices and are challenging to compute. We explicitly compute diagram 5 (e). We construct this diagram using Eq.\eqref{emscalar} and Eq.\eqref{fig:3pointv1}, then appropriately contract the NRG fields using Eq.\eqref{scalargraviton} and the EM fields using Eq.\eqref{emscalarprop}. Integrating over time, $d^3x$, and $d^3y$ further simplifies the integral, resulting in:

\begin{widetext}
\begin{equation}
\begin{split}
	\mathrm{Diagram \; } \ref{fig:Gq^4} \mathrm{(e)}
	&= \alpha \int dt \int_{{\bm k_1}, {\bm k_2}, {\bm k_3}} e^{i({\bm k}_1+{\bm k}_3)\cdot {\bm r}}\frac{({\bm k}_1\cdot{\bm k}_2) ({\bm k}_3\cdot (-{\bm k}_1-{\bm k}_2-{\bm k}_3)}{{\bm k}_1^2{\bm k}_2^2({\bm k}_1+{\bm k}_2)^2{\bm k}_3^2({\bm k}_1+{\bm k}_2+{\bm k}_3)^2},
\end{split}
\end{equation}
\end{widetext}
where $\alpha = q_1^2q_2^2 (4\pi G) (-4\pi)^4/(-2\pi)^2$. To simplify further, we redefine the momenta as ${\bm k_1} \rightarrow {\bm k_1} + {\bm k_3}$ and ${\bm k_2} \rightarrow {\bm k_2} + {\bm k_1}$:
\begin{widetext}
\begin{equation}
\begin{split}
	\mathrm{Diagram \; } \ref{fig:Gq^4} \mathrm{(e)}
	&= \alpha \int dt \int_{{\bm k_1}, {\bm k_2}, {\bm k_3}} e^{i{\bm k_1}\cdot {\bm r}}\frac{({\bm k_1}-{\bm k_3})\cdot({\bm k_2}-{\bm k_1}) (-{\bm k_3}\cdot{\bm k_2})}{({\bm k_1}-{\bm k_3})^2({\bm k_2}-{\bm k_1})^2({\bm k_2}-{\bm k_3})^2{\bm k_3}^2{\bm k_2}^2}.
\end{split}
\end{equation}
\end{widetext}
This integral is irreducible. We further split the terms one by one and solve them individually.
\begin{equation}
\begin{split}
	\mathrm{Diagram \; } \ref{fig:Gq^4} \mathrm{(e)}
	&= \alpha \int dt \int_{\bm k_1} e^{i{\bm k_1}\cdot {\bm r}}(\mathbf{I}_1 + \mathbf{I}_2 + \mathbf{I}_3 + \mathbf{I}_4),
\end{split}
\end{equation}
where the terms $\mathbf{I}_1$, $\mathbf{I}_2$, $\mathbf{I}_3$, and $\mathbf{I}_4$ are defined as
\begin{eqnarray}
    \mathbf{I}_1 =  \int_{k_2, k_3} \frac{({\bm k_1}\cdot{\bm k_2}) (-{\bm k_3}\cdot{\bm k_2})}{({\bm k_1}-{\bm k_3})^2({\bm k_2}-{\bm k_1})^2({\bm k_2}-{\bm k_3})^2{\bm k_3}^2{\bm k_2}^2},\nonumber\\
    \mathbf{I}_2 =  \int_{k_2, k_3} \frac{(-{\bm k_1}\cdot{\bm k_1}) (-{\bm k_3}\cdot{\bm k_2})}{({\bm k_1}-{\bm k_3})^2({\bm k_2}-{\bm k_1})^2({\bm k_2}-{\bm k_3})^2{\bm k_3}^2{\bm k_2}^2},\nonumber\\
    \mathbf{I}_3 = \int_{k_2, k_3} \frac{(-{\bm k_3}\cdot{\bm k_2}) (-{\bm k_3}\cdot{\bm k_2})}{({\bm k_1}-{\bm k_3})^2({\bm k_2}-{\bm k_1})^2({\bm k_2}-{\bm k_3})^2{\bm k_3}^2{\bm k_2}^2},\nonumber\\
    \mathbf{I}_4 = \int_{k_2, k_3} \frac{({\bm k_3}\cdot{\bm k_1}) (-{\bm k_3}\cdot{\bm k_2})}{({\bm k_1}-{\bm k_3})^2({\bm k_2}-{\bm k_1})^2({\bm k_2}-{\bm k_3})^2{\bm k_3}^2{\bm k_2}^2}\nonumber.
\end{eqnarray}
These integrals can be solved using the tensor two-loop integral, Eq. \eqref{eq:2tensor2loop}. While performing this integral, di- vergences are managed using regularization techniques. Specifically, we compute the integral in $3 - \epsilon$ dimensions and then take the limit as $\epsilon \rightarrow 0$, yielding a finite result. The final result:

\begin{flalign}
\begin{split}
	\mathrm{Diagram \; } \ref{fig:Gq^4} \mathrm{(e)}
	&=  Gq_1^2q_2^2\int d t \frac{1}{r^3}.
\end{split}
\end{flalign}

Similarly, 
\begin{flalign}
\begin{split}
	\mathrm{Diagram \; } \ref{fig:Gq^4} \mathrm{(a)} &= -\frac{1}{3}Gq_1q_2^3 \int d t \frac{1}{r^3},
\end{split}
\end{flalign}

\begin{flalign}
\begin{split}
	\mathrm{Diagram \; } \ref{fig:Gq^4} \mathrm{(b)}
	&= \frac{1}{3}Gq_1q_2^3\int d t \frac{1}{r^3},
\end{split}
\end{flalign}

The factorizable 2-loop diagrams 5(f) and 5(g) produce short-distance contributions that ultimately cancel out:
\begin{flalign}
\begin{split}
	\mathrm{Diagram \; } \ref{fig:Gq^4} \mathrm{(c)}
	&= 0,
\end{split}
\end{flalign}

\begin{flalign}
\begin{split}
	\mathrm{Diagram \; } \ref{fig:Gq^4} \mathrm{(d)} &= 0,
\end{split}
\end{flalign}

\begin{flalign}
\begin{split}
	\mathrm{Diagram \; } \ref{fig:Gq^4} \mathrm{(f)}
	&= 2Gq_1^2q_2^2\int d t \frac{1}{r^3}.
\end{split}
\end{flalign}

\subsection{$G^2q^2$ order diagrams}
\label{subsection:G^2q^2}
\begin{figure*}
\centering
   \includegraphics{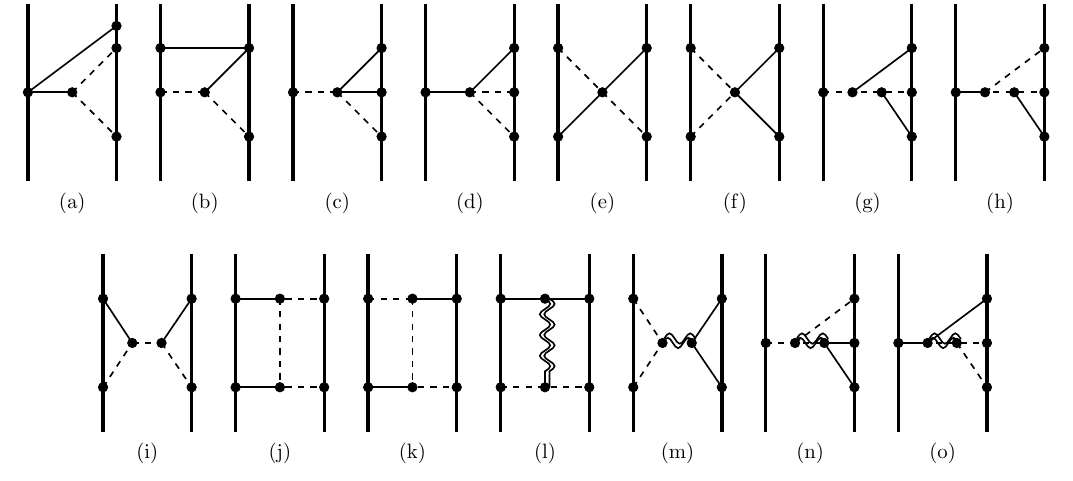} 
\caption{Diagrams that have $\mathcal{O}(G^2q^2)$.} 
\label{fig:G^2q^2}  
\end{figure*}
At the $\mathcal{O}(G^2q^2)$ order, there are 15 two-loop diagrams. The diagrams 6(c), 6(d), 6(g), 6(h), 6(n) and 6(o) are nested 2-loop types, while 6(e), 6(f), 6(i) and 6(m) are factorizable 2-loop types, and 6(j), 6(k) and 6(l) are irreducible 2-loop types. The symmetry factors of diagrams 6(a), 6(c), 6(d), 6(j), 6(n) and 6(o) are $1/2$, for diagrams 6(e) and 6(f) it is $1/4$, and the remaining diagrams have a symmetry factor of 1.  The diagrams 6(a) and 6(b) contain one 3-point vertex and are solved using the 1-loop master integral (\ref{eq:1loop}). The results are:








\begin{flalign}
\begin{split}
	\mathrm{Diagram \; } \ref{fig:G^2q^2} \mathrm{(a)}
	&= G^2q_2^2m_1m_2 \int d t \frac{1}{2r^3}, 
	\end{split}
\end{flalign}

\begin{flalign}
\begin{split}
	\mathrm{Diagram \; } \ref{fig:G^2q^2} \mathrm{(b)}
	&= -G^2q_1q_2m_1m_2\int d t \frac{1}{r^3},
\end{split}
\end{flalign}



The diagrams 6(c), 6(d), 6(e), and 6(f) involve a single 4-point vertex. We demonstrate the computation for diagram 6(c). Constructing the diagram as usual, using Eq.\eqref{emscalar}, Eq.\eqref{gravitonscalarvertex}, and Eq.~\eqref{4pointvertex}, perform the integration over $d^3x$ and time, we get:

\begin{equation}
\begin{split}
	\mathrm{Diagram \; } \ref{fig:G^2q^2} \mathrm{(c)}
	&= G^2q_1q_2m_2^2\frac{1}{\pi}(-4\pi)^2\int dt \frac{1}{r^2} \int_{\bm k_1} \frac{e^{i {\bm k_1}\cdot {\bm r}}}{{\bm k_1}^2}.
\end{split}
\end{equation}
Carrying out the Fourier integral using Eq.~\eqref{eq:ft}, we obtain:
\begin{flalign}
\begin{split}
	\mathrm{Diagram \; } \ref{fig:G^2q^2} \mathrm{(c)}
	&= \frac{2}{3}G^2q_1q_2m_2^2 \int d t \frac{1}{r^3}.
\end{split}
\end{flalign}
The remaining diagrams, 6(d), 6(e), and 6(l), are evaluated in a similar manner. The results are:
\begin{flalign}
\begin{split}
	\mathrm{Diagram \; } \ref{fig:G^2q^2} \mathrm{(d)}
	&= -\frac{1}{3}G^2q_2^2m_1m_2 \int d t \frac{1}{r^3},
\end{split}
\end{flalign}
\begin{flalign}
\begin{split}
	\mathrm{Diagram \; } \ref{fig:G^2q^2} \mathrm{(e)}
	&= 0,
\end{split}
\end{flalign}

\begin{flalign}
\begin{split}
	\mathrm{Diagram \; } \ref{fig:G^2q^2} \mathrm{(f)}
	&= 0.
\end{split}
\end{flalign}

The remaining diagrams involve two 3-point vertices and are calculated similarly to diagram \ref{fig:Gq^4}(j), yielding the following result:
\begin{flalign}
\begin{split}
	\mathrm{Diagram \; } \ref{fig:G^2q^2} \mathrm{(g)}
	&= -\frac{4}{3}G^2q_1q_2m_2^2 \int d t \frac{1}{r^3},
\end{split}
\end{flalign}

\begin{flalign}
\begin{split}
	\mathrm{Diagram \; } \ref{fig:G^2q^2} \mathrm{(h)}
	&= \frac{2}{3}G^2m_1m_2q_2^2 \int d t \frac{1}{r^3},
\end{split}
\end{flalign}

\begin{flalign}
\begin{split}
	\mathrm{Diagram \; } \ref{fig:G^2q^2} \mathrm{(i)}
	&= 0,
\end{split}
\end{flalign}

\begin{flalign}
\begin{split}
	\mathrm{Diagram \; } \ref{fig:G^2q^2} \mathrm{(j)}
	&= G^2q_2^2m_1^2 \int d t \frac{1}{r^3},
\end{split}
\end{flalign}

\begin{flalign}
\begin{split}
	\mathrm{Diagram \; } \ref{fig:G^2q^2} \mathrm{(k)}
	&= -14G^2q_1q_2m_1m_2 \int d t \frac{1}{r^3}.
\end{split}
\end{flalign}

\begin{flalign}
\begin{split}
	\mathrm{Diagram \; } \ref{fig:G^2q^2} \mathrm{(l)}
	&= -4G^2m_1m_2q_1q_2 \int d t \frac{1}{r^3},
\end{split}
\end{flalign}

\begin{flalign}
\begin{split}
	\mathrm{Diagram \; } \ref{fig:G^2q^2} \mathrm{(m)}
	&= 0,
\end{split}
\end{flalign}

\begin{flalign}
\begin{split}
	\mathrm{Diagram \; } \ref{fig:G^2q^2} \mathrm{(n)}
	&= -\frac{1}{3}G^2q_1q_2m_2^2 \int d t \frac{1}{r^3},
\end{split}
\end{flalign}

\begin{flalign}
\begin{split}
	\mathrm{Diagram \; } \ref{fig:G^2q^2} \mathrm{(o)}
	&= -\frac{1}{3}G^2q_2^2m_1m_2 \int d t \frac{1}{r^3}.
\end{split}
\end{flalign}

\section{Results}
\label{sec:results}
Having computed all the diagrams, we can now construct the 2PN interaction Lagrangian. Two main contributions are required: the kinetic energy and the potential from the diagrams. The kinetic energy is obtained by expanding the matter coupling action, Eq.~(\ref{eq:sppkk}), to $\mathcal{O}(v^6)$ while setting all fields to zero. The final potential is derived by summing the potential contributions from each diagram.
Adding the contributions from all 2PN diagrams and including the kinetic energy term. We obtain the  Lagrangian at 2PN order
\begin{widetext}
    \begin{align}
        \mathcal{L}_{2PN} &= \frac{1}{16}m_1v_1^6 + \frac{G}{r^2} \left[ m_1 \left( 
                 \frac{1}{2} q_2^2 ({\bm v}_1 \cdot {\bm v}_2)
                 - \frac{1}{4} q_2^2 ({\bm v}_1 \cdot {\bm v}_1) 
                -\frac{1}{2} q_1 q_2 ({\bm v}_1 \cdot {\bm v}_1)   
            \right) \right. \nonumber \\ 
        & \left. \quad + m_1 \left( 
                q_1 q_2 ({\bm v}_1 \cdot {\bm n})^2 
                - \frac{1}{2} q_2^2 ({\bm v}_1 \cdot {\bm n})^2 
                - 2 q_1 q_2 ({\bm v}_1 \cdot {\bm n}) ({\bm v}_2 \cdot {\bm n}) 
                + 2 q_2^2 ({\bm v}_1 \cdot {\bm n}) ({\bm v}_2 \cdot {\bm n}) 
                - q_2^2 ({\bm v}_2 \cdot {\bm n})^2 
            \right) 
        \right] \nonumber \\& \quad
        + \frac{3}{2}Gq_1^2q_2^2 \frac{1}{r^3} 
        + \frac{G^2}{r^3} \left(  m_1 m_2 \left( 
                \frac{1}{2}q_1^2 - 10 q_1 q_2 
            \right) 
            + m_1^2 (q_2^2 - q_1 q_2) 
        \right) \nonumber\\& \quad  + q_1q_2r\left(\frac{1}{16}({\bm a}_1 \cdot {\bm n}) ({\bm a}_2 \cdot {\bm n})  - \frac{3}{16} ({\bm a}_1 \cdot {\bm a}_2)\right)
        +  q_1q_2 \left(\frac{1}{4} ({\bm a}_1 \cdot {\bm v}_2) ({\bm v}_2 \cdot {\bm n}) + \frac{1}{8} ({\bm v}_1 \cdot {\bm v}_1) ({\bm a}_2 \cdot {\bm n}) + \frac{1}{8} ({\bm a}_1 \cdot {\bm n}) ({\bm v}_2 \cdot {\bm n})^2 \right) \nonumber \\ 
        & \quad
        +\frac{q_1 q_2}{r} \left( 
        \frac{1}{8} ({\bm v}_1 \cdot {\bm v}_2)^2 - \frac{1}{16} ({\bm v}_1 \cdot {\bm v}_1) ({\bm v}_2 \cdot {\bm v}_2) 
        - \frac{3}{16} ({\bm v}_1 \cdot {\bm n})^2 ({\bm v}_2 \cdot {\bm n})^2 + \frac{1}{8} ({\bm v}_1 \cdot {\bm v}_1) ({\bm v}_2 \cdot {\bm n})^2 \right) + (1 \leftrightarrow 2)
    \end{align}
\end{widetext}
where we need to include the contributions from interchanging $1 \leftrightarrow 2$  and $\bm{n} \rightarrow -\bm{n}$, resulting in different diagrams. This result agrees with the calculation done with the EFTofPNG package \cite{Levi:2017kzq}. In the $G\rightarrow 0$ limit (purely electromagnetic two-body dynamics), this Lagrangian agrees with the known 2PN result \cite{Barker:1980kef}.

\section{Conclusions}
\label{sec:conclusions}
We have outlined using the EFT method to efficiently calculate the 2PN order Lagrangian of Einstein-Maxwell theory for a binary system. We derived the Feynman rules up to 2PN order and depicted the 1PN and 2PN order Feynman diagrams contributing to the Lagrangian. The calculation included $\mathcal{O}(q^2v^4)$, $\mathcal{O}(Gq^2v^2)$, $\mathcal{O}(G^2q^2)$, and $\mathcal{O}(Gq^4)$ types of 2PN order diagrams, encompassing 1-loop and 2-loop integrals. All diagrams could be computed using just two master integrals, showcasing the efficiency of the EFT method for calculating the 2PN Lagrangian.

For the Einstein–Maxwell system, dissipative far-zone effects enter earlier than in pure GR: electromagnetic dipole radiation induces radiation reaction at 1.5PN \cite{Jackson:1998}, whereas the leading gravitational-wave reaction appears at 2.5PN \cite{Blanchet:2014LRR}. Conservative hereditary (tail) effects first arise at 4PN from gravitational-wave tails \cite{Foffa:2013Tail,Blanchet:2014LRR}; in flat spacetime the Maxwell field satisfies Huygens’ principle, so EM tails are absent, implying any mixed gravity-EM conservative hereditary terms occur no earlier than 4PN \cite{Harte:2013NoEMTails}. Finite-size (tidal) corrections from electromagnetic polarizability contribute as \(\sim \alpha_E E^2 \sim \alpha_E q^2/r^4\), which corresponds to 3PN in standard PN counting for \(\alpha_E\sim R^3\) \cite{Jackson:1998,GoldbergerRothstein:2006EFT}, while the leading gravitational quadrupolar tides enter at 5PN \cite{FlanaganHinderer:2008PRD,Hinderer:2008ApJ}. All of these sectors lie beyond the 2PN conservative near-zone order computed here and will be addressed in future work.


An obvious extension of the current result is to compute the 3PN Lagrangian, which will involve up to 3-loop Feynman integrals. The number of diagrams to be evaluated is large, and we leave this for future work. Finally, with the Lagrangian in hand, the next step will be to calculate the orbital motion and the gravitational radiation emitted; this is work in progress.



\acknowledgements

The author is very grateful to Jan Steinhoff for suggesting EFT method and valuable discussions throughout the project. The author would also like to thank Chris Van Den Broeck for going through the manuscript. The author thanks Tanja Hinderer and Jasper Roosmale Nepveu for valuable comments to improve the manuscript.
This work was supported by the Netherlands Organization for Scientific Research (NWO).

\appendix

\section{Useful formulas}
\label{sec:appendix}
We flip the time derivative between the two particles  by using the identity
\begin{widetext}
\begin{align}
    \label{eq:timeflip}
    \int dt_1 dt_2  \partial_{t_1}\delta(t_1-t_2) f(t_1) g(t_2) 
    &= -\int dt_1  dt_2  \partial_{t_2}\delta(t_1-t_2) f(t_1) g(t_2).
\end{align}
\end{widetext}
We evaluate the Fourier integrals that we often encounter from propagators and loop integrals using the d-dimensional master formula given by 
\begin{equation}
\label{eq:ft}
\int \frac{d^d \mathbf{k}}{(2\pi)^d} \frac{e^{i \mathbf{k} \cdot \mathbf{r}}}{(\mathbf{k}^2)^\alpha}
= \frac{1}{(4\pi)^{d/2}} \frac{\Gamma(d/2 - \alpha)}{\Gamma(\alpha)} 
\left(\frac{\mathbf{r}^2}{4}\right)^{\alpha - d/2}.
\end{equation}
From this master formula, we obtain the following required Fourier integrals \cite{Levi:2011eq}:
\begin{widetext}
\begin{eqnarray}
\label{tensorfourierindentity}
I^i & \equiv & \int \frac{d^d \mathbf{k}}{(2\pi)^d} \frac{k^i e^{i \mathbf{k} \cdot \mathbf{r}}}{(\mathbf{k}^2)^\alpha}  =  \frac{i}{(4\pi)^{d/2}} \frac{\Gamma(d/2 - \alpha + 1)}{\Gamma(\alpha)}
\left(\frac{\mathbf{r}^2}{4}\right)^{\alpha - d/2 - 1/2} n^i, \\
I^{ij} & \equiv & \int \frac{d^d \mathbf{k}}{(2\pi)^d} \frac{k^i k^j e^{i \mathbf{k} \cdot \mathbf{r}}}{(\mathbf{k}^2)^\alpha} = \frac{1}{(4\pi)^{d/2}} \frac{\Gamma(d/2 - \alpha + 1)}{\Gamma(\alpha)}
\left(\frac{\mathbf{r}^2}{4}\right)^{\alpha - d/2 - 1}
\left(\frac{1}{2} \delta^{ij} + (\alpha - 1 - d/2) n^i n^j \right), \\
I^{ijl} & \equiv & \int \frac{d^d \mathbf{k}}{(2\pi)^d} \frac{k^i k^j k^l e^{i \mathbf{k} \cdot \mathbf{r}}}{(\mathbf{k}^2)^\alpha} = \frac{i}{(4\pi)^{d/2}} \frac{\Gamma(d/2 - \alpha + 2)}{\Gamma(\alpha)}
\left(\frac{\mathbf{r}^2}{4}\right)^{\alpha - d/2 - 3/2} \nonumber \\
& & \times \left( \frac{1}{2} \left(\delta^{ij} n^l + \delta^{il} n^j + \delta^{jl} n^i \right)
+ (\alpha - d/2 - 2) n^i n^j n^l \right), \\
I^{ijlm} & \equiv & \int \frac{d^d \mathbf{k}}{(2\pi)^d} \frac{k^i k^j k^l k^m e^{i \mathbf{k} \cdot \mathbf{r}}}{(\mathbf{k}^2)^\alpha} = \frac{1}{(4\pi)^{d/2}} \frac{\Gamma(d/2 - \alpha + 2)}{\Gamma(\alpha)}
\left(\frac{\mathbf{r}^2}{4}\right)^{\alpha - d/2 - 2}  \times \left( \frac{1}{4} \left( \delta^{ij} \delta^{lm} + \delta^{il} \delta^{jm} + \delta^{im} \delta^{jl} \right) \right. \nonumber \\
& & + \frac{\alpha - d/2 - 2}{2} \left( \delta^{ij} n^l n^m + \delta^{il} n^j n^m 
+ \delta^{im} n^j n^l + \delta^{jl} n^i n^m + \delta^{jm} n^i n^l + \delta^{lm} n^i n^j \right) \nonumber \\
& & \left. + (\alpha - d/2 - 2)(\alpha - d/2 - 3) n^i n^j n^l n^m \right).
\end{eqnarray}
\end{widetext}

We use the d-dimensional master formula for one-loop scalar integrals given by 
\begin{widetext}
\begin{equation}
    J \equiv \int \frac{d^d \mathbf{k}}{(2\pi)^d} \frac{1}{\left[ \mathbf{k}^2 \right]^\alpha \left[ (\mathbf{k} - \mathbf{q})^2 \right]^\beta} 
    = \frac{1}{(4\pi)^{d/2}} \frac{\Gamma(\alpha + \beta - d/2)}{\Gamma(\alpha) \Gamma(\beta)} 
    \frac{\Gamma(d/2 - \alpha) \Gamma(d/2 - \beta)}{\Gamma(d - \alpha - \beta)} 
    \left( q^2 \right)^{d/2 - \alpha - \beta}.
    \label{eq:1loop}
\end{equation}
\end{widetext}
The d-dimensional master formula for one-loop tensor integrals is also taken from \cite{Levi:2011eq}. 
Similarly, one can also derive the following d-dimensional master formulae for the one-loop tensor integrals: 
\begin{widetext}
\begin{align}
    \label{eq:1loopvec}
    J^i & \equiv \int \frac{d^d \mathbf{k}}{(2\pi)^d} \frac{k^i}{\left[\mathbf{k}^2\right]^\alpha \left[(\mathbf{k} - \mathbf{q})^2\right]^\beta} = \frac{1}{(4\pi)^{d/2}} \frac{\Gamma(\alpha + \beta - d/2)}{\Gamma(\alpha) \Gamma(\beta)} 
    \frac{\Gamma(d/2 - \alpha + 1) \Gamma(d/2 - \beta)}{\Gamma(d - \alpha - \beta + 1)} 
    \left(q^2\right)^{d/2 - \alpha - \beta} q^i, \\
    J^{ij} & \equiv \int \frac{d^d \mathbf{k}}{(2\pi)^d} \frac{k^i k^j}{\left[\mathbf{k}^2\right]^\alpha \left[(\mathbf{k} - \mathbf{q})^2\right]^\beta} = \frac{1}{(4\pi)^{d/2}} \frac{\Gamma(\alpha + \beta - d/2 - 1)}{\Gamma(\alpha) \Gamma(\beta)} 
    \frac{\Gamma(d/2 - \alpha + 1) \Gamma(d/2 - \beta)}{\Gamma(d - \alpha - \beta + 2)} 
    \left(q^2\right)^{d/2 - \alpha - \beta} \nonumber \\
    & \quad \times \left[ \frac{d/2 - \beta}{2} q^2 \delta^{ij} + (\alpha + \beta - d/2 - 1) (d/2 - \alpha + 1) q^i q^j \right], \\
    J^{ijl} & \equiv \int \frac{d^d \mathbf{k}}{(2\pi)^d} \frac{k^i k^j k^l}{\left[\mathbf{k}^2\right]^\alpha \left[(\mathbf{k} - \mathbf{q})^2\right]^\beta} = \frac{1}{(4\pi)^{d/2}} \frac{\Gamma(\alpha + \beta - d/2 - 1)}{\Gamma(\alpha) \Gamma(\beta)} 
    \frac{\Gamma(d/2 - \alpha + 2) \Gamma(d/2 - \beta)}{\Gamma(d - \alpha - \beta + 3)} 
    \left(q^2\right)^{d/2 - \alpha - \beta} \nonumber \\
    & \quad \times \left[ \frac{d/2 - \beta}{2} q^2 \left(\delta^{ij} q^l + \delta^{il} q^j + \delta^{jl} q^i \right) + (\alpha + \beta - d/2 - 1) (d/2 - \alpha + 2) \frac{d/2 - \beta}{2} q^2 \right. \nonumber \\
    & \quad \times \left(\delta^{ij} q^l q^m + \delta^{il} q^j q^m + \delta^{im} q^j q^l + \delta^{jl} q^i q^m + \delta^{jm} q^i q^l + \delta^{lm} q^i q^j \right) \nonumber \\
    & \quad \left. + (\alpha + \beta - d/2 - 2) (\alpha + \beta - d/2 - 1) (d/2 - \alpha + 2) (d/2 - \alpha + 3) q^i q^j q^l q^m \right].
\end{align}
\end{widetext}

In addition, we encounter irreducible two-loop tensor integrals up to order 4. Using an integration by parts method as in \cite{Smirnov:2004ym}, these can be written as a sum of factorizable and nested two-loops, as explained in Sec.~\ref{subsection:Gq^4},\ref{subsection:G^2q^2}. The required irreducible two-loop tensor integral reductions are given by
\begin{widetext}
\begin{align}
    \label{eq:2tensor2loop}
    \int_{\mathbf{k}_1 \mathbf{k}_2} \frac{k_1^i k_2^j}{\bm{k}_1^2 \left(p - k_1\right)^2 \bm{k}_2^2 \left(p - k_2\right)^2 \left(k_1 - k_2\right)^2} 
    &= \frac{1}{d - 3} \int_{\mathbf{k}_1 \mathbf{k}_2} \Bigg[ 
        \frac{p^i k_2^j}{k_1^4 \left(p - k_1\right)^2 \bm{k}_2^2 \left(p - k_2\right)^2} 
        - \frac{k_1^i k_2^j}{k_1^4 \left(p - k_1\right)^2 \left(p - k_2\right)^2 \left(k_1 - k_2\right)^2} \nonumber \\
    & \quad - \frac{k_1^i k_2^j}{\bm{k}_1^2 \left(p - k_1\right)^4 \bm{k}_2^2 \left(k_1 - k_2\right)^2} 
        + \frac{1}{d - 4} \Bigg( 
            2 \frac{k_2^i k_2^j}{k_1^4 \left(p - k_1\right)^2 \bm{k}_2^2 \left(p - k_2\right)^2} 
            \nonumber \\
    & \quad - \frac{k_2^i k_2^j}{k_1^4 \left(p - k_1\right)^2 \left(p - k_2\right)^2 \left(k_1 - k_2\right)^2}  - \frac{k_2^i k_2^j}{\bm{k}_1^2 \left(p - k_1\right)^4 \bm{k}_2^2 \left(k_1 - k_2\right)^2} 
        \Bigg) \Bigg], \\
    \label{eq:3tensor2loop}
    \int_{\mathbf{k}_1 \mathbf{k}_2} \frac{k_1^i k_1^j k_2^l}{\bm{k}_1^2 \left(p - k_1\right)^2 \bm{k}_2^2 \left(p - k_2\right)^2 \left(k_1 - k_2\right)^2} 
    &= \frac{1}{d - 3} \int_{\mathbf{k}_1 \mathbf{k}_2} \Bigg[ 
        \frac{p^l k_1^i k_1^j}{\bm{k}_1^2 \left(p - k_1\right)^2 k_2^4 \left(p - k_2\right)^2} 
        - \frac{k_1^i k_1^j k_2^l}{\left(p - k_1\right)^2 k_2^4 \left(p - k_2\right)^2 \left(k_1 - k_2\right)^2} \nonumber \\
    & \quad - \frac{k_1^i k_1^j k_2^l}{\bm{k}_1^2 \bm{k}_2^2 \left(p - k_2\right)^4 \left(k_1 - k_2\right)^2} 
        + \frac{1}{d - 4} \Bigg( 
            2 \frac{k_1^i k_1^j k_1^l}{\bm{k}_1^2 \left(p - k_1\right)^2 k_2^4 \left(p - k_2\right)^2} \nonumber \\
    & \quad
            - \frac{k_1^i k_1^j k_1^l}{\left(p - k_1\right)^2 k_2^4 \left(p - k_2\right)^2 \left(k_1 - k_2\right)^2}  - \frac{k_1^i k_1^j k_1^l}{\bm{k}_1^2 \bm{k}_2^2 \left(p - k_2\right)^4 \left(k_1 - k_2\right)^2} 
        \Bigg) \Bigg], \\
    \label{eq:4tensor2loop}
    \int_{\mathbf{k}_1 \mathbf{k}_2} \frac{k_1^i k_1^j k_2^l k_2^m}{\bm{k}_1^2 \left(p - k_1\right)^2 \bm{k}_2^2 \left(p - k_2\right)^2 \left(k_1 - k_2\right)^2} 
    &= \frac{1}{d - 2} \int_{\mathbf{k}_1 \mathbf{k}_2} \Bigg[ 
        \frac{k_1^i k_1^j k_2^l k_2^m}{k_1^4 \left(p - k_1\right)^2 \bm{k}_2^2 \left(p - k_2\right)^2} 
        + \frac{k_1^i k_1^j k_2^l k_2^m}{\bm{k}_1^2 \left(p - k_1\right)^4 \bm{k}_2^2 \left(p - k_2\right)^2} \nonumber \\
    & \quad - \frac{k_1^i k_1^j k_2^l k_2^m}{k_1^4 \left(p - k_1\right)^2 \left(p - k_2\right)^2 \left(k_1 - k_2\right)^2} 
        - \frac{k_1^i k_1^j k_2^l k_2^m}{\bm{k}_1^2 \left(p - k_1\right)^4 \bm{k}_2^2 \left(k_1 - k_2\right)^2} \nonumber \\
    & \quad + \frac{1}{d - 3} \Bigg( 
            \frac{p^i k_2^j k_2^l k_2^m}{k_1^4 \left(p - k_1\right)^2 \bm{k}_2^2 \left(p - k_2\right)^2} 
            + \frac{p^j k_2^i k_2^l k_2^m}{k_1^4 \left(p - k_1\right)^2 \bm{k}_2^2 \left(p - k_2\right)^2} \nonumber \\
    & \quad - \frac{k_1^i k_2^j k_2^l k_2^m}{k_1^4 \left(p - k_1\right)^2 \left(p - k_2\right)^2 \left(k_1 - k_2\right)^2} 
            - \frac{k_1^j k_2^i k_2^l k_2^m}{k_1^4 \left(p - k_1\right)^2 \left(p - k_2\right)^2 \left(k_1 - k_2\right)^2} \nonumber \\
    & \quad - \frac{k_1^i k_2^j k_2^l k_2^m}{\bm{k}_1^2 \left(p - k_1\right)^4 \bm{k}_2^2 \left(k_1 - k_2\right)^2} 
            - \frac{k_1^j k_2^i k_2^l k_2^m}{\bm{k}_1^2 \left(p - k_1\right)^4 \bm{k}_2^2 \left(k_1 - k_2\right)^2} \nonumber \\
    & \quad + \frac{2}{d - 4} \Bigg( 
            2 \frac{k_2^i k_2^j k_2^l k_2^m}{k_1^4 \left(p - k_1\right)^2 \bm{k}_2^2 \left(p - k_2\right)^2} 
            - \frac{k_2^i k_2^j k_2^l k_2^m}{k_1^4 \left(p - k_1\right)^2 \left(p - k_2\right)^2 \left(k_1 - k_2\right)^2} \nonumber \\
& \quad - \frac{k_2^i k_2^j k_2^l k_2^m}{\bm{k}_1^2 \left(p - k_1\right)^4 \bm{k}_2^2 \left(k_1 - k_2\right)^2}
\Bigg) \Bigg].
\end{align}
\end{widetext}
It should be noted that these expressions contain explicit poles in $d=3$, but these cancel out in the dimensional regularization. 

\bibliography{main_auto}

\end{document}